\documentclass[
aps, 
twocolumn, 
nofootinbib]{revtex4-1}

\usepackage{graphicx}
\usepackage{dcolumn}
\usepackage{bm}
\usepackage{amsmath, amssymb, amsthm}
\usepackage{color}
\usepackage{adjustbox}
\usepackage[utf8]{inputenc}
\usepackage[english]{babel}
\usepackage{comment}
\usepackage{algorithm,algpseudocode}
\usepackage{booktabs}
\usepackage[caption=false]{subfig} 

\DeclareFontFamily{U}{mathx}{\hyphenchar\font45}
\DeclareFontShape{U}{mathx}{m}{n}{<-> mathx10}{}
\DeclareSymbolFont{mathx}{U}{mathx}{m}{n}
\DeclareMathAccent{\widebar}{0}{mathx}{"73}

\expandafter\def\expandafter\normalsize\expandafter{%
  \normalsize  
  \setlength\abovedisplayskip{3ex}
  \setlength\belowdisplayskip{3ex}
  \setlength\abovedisplayshortskip{3ex}
  \setlength\belowdisplayshortskip{3ex}
}

\newcommand{\papertitle}{Modeling the Impact of Social Distancing and Targeted Vaccination on the Spread of COVID-19 through a Real City-Scale Contact Network}

\begin{document}


\title{\papertitle}

\author{Gavin S. Hartnett\footnote{Corresponding author: \texttt{hartnett@rand.org}}}
\author{Edward~Parker} 
\author{Timothy~R.~Gulden}
\author{\\Raffaele~Vardavas}
\author{David~Kravitz}
\affiliation{
\vskip 1.5em
RAND Corporation \\
1776 Main St, Santa Monica, CA 90401 
}


\begin{abstract}
We use mobile device data to construct empirical interpersonal physical contact networks in the city of Portland, Oregon, both before and after social distancing measures were enacted during the COVID-19 pandemic. These networks reveal how social distancing measures and the public's reaction to the incipient pandemic affected the connectivity patterns within the city. We find that as the pandemic developed there was a substantial decrease in the number of individuals with many contacts. We further study the impact of these different network topologies on the spread of COVID-19 by simulating an SEIR epidemic model over these networks, and find that the reduced connectivity greatly suppressed the epidemic. We then investigate how the epidemic responds when part of the population is vaccinated, and we compare two vaccination distribution strategies, both with and without social distancing. Our main result is that the heavy-tailed degree distribution of the contact networks causes a targeted vaccination strategy that prioritizes high-contact individuals to reduce the number of cases far more effectively than a strategy that vaccinates individuals at random. Combining both targeted vaccination and social distancing leads to the greatest reduction in cases, and we also find that the marginal benefit of a targeted strategy as compared to a random strategy exceeds the marginal benefit of social distancing for reducing the number of cases. These results have important implications for ongoing vaccine distribution efforts worldwide.
\end{abstract}

\maketitle

\section{Introduction}
Real-world epidemics play out over person-to-person physical contact networks. In order for an infectious disease to spread from one person to another, those people need to come into either direct or indirect physical contact. For airborne diseases such as COVID-19, transmissions are most often due to contemporaneous, close-proximity physical interactions, with transmission especially likely if neither party is wearing a mask and if the interaction occurs indoors over an extended period of time. For diseases such as HIV/AIDS, the physical contact could be due to a sexual interaction or the sharing of needles, while still other diseases may be transmitted by a shared contact with an infected surface. In each case the transmission is facilitated through a physical connection of some sort, which may or may not involve the two individuals occupying the same place at the same time.  

Consequently, the disease can be thought of as spreading across a network or graph $G$, with the nodes representing individuals and the edges representing physical contacts that can facilitate the transmission of the infectious agent. This network is incredibly important for the evolution of the epidemic. For example, the basic reproductive number $R_0$, defined as the expected number of infections caused by an infected individual in a uniformly susceptible population, is a key statistic that has been heavily used by policy makers and scientists in battling the COVID-19 epidemic. $R_0$ depends on both biological properties of the disease, for example on how infectious it is or how long an infected individual is contagious, as well as on the behavior of the public. Many of these behavioral properties, such as the extent to which social distancing measures are being observed, are captured in the statistical properties of the contact network. As a result, $R_0$ is a function of the underlying contact network: the same disease could spread qualitatively differently through two separate populations with different contact networks.
It is therefore extremely important to both estimate $G$ and incorporate that estimate in epidemiological modeling efforts. 

Motivated by the ongoing COVID-19 epidemic, there have been several attempts to estimate or infer the contact network $G$ using mobile device or cell-phone data. In many countries, most individuals own smartphones with location-tracking apps installed, and these individuals are highly likely to keep their phone with them when they leave their home and interact with people outside of their household. These mobile device datasets may also be used to measure intra-household contacts by noting whether two devices are frequently co-located during evenings. These cell-phone data can be used to estimate a real-world contact network that, in principle, can capture a large fraction of all contacts within a population. Of course, there are important details of a physical connection that cannot be inferred from mobile device data alone, such as whether either party was wearing a mask, but nonetheless such mobile device-derived contact networks represent a significant resource for epidemiologists and public policy makers \cite{grantz2020use, oliver2020mobile}. In particular, this data may be used to study the compliance with and impact of social distancing guidelines \cite{heiler2020country, schlosser2020covid}. In addition to providing insight into the mobility patterns of a population, these datasets may be used to construct more sophisticated epidemiological models -- for example, capturing local variations in connectivity and modeling the impact across demographic groups by using mobility data to model the disease at the census block level \cite{chang2021mobility}.

As the pandemic enters its second year and global vaccination efforts intensify, epidemiologists have turned to these models to study vaccine distribution strategies. It was known prior to COVID-19 that a population's connectivity patterns play an important role in the impact of a vaccination strategy, especially across different demographic groups \cite{medlock2009optimizing}. This information was previously incorporated into models using very coarse-grained mixing matrices, but real-world contact networks can capture more fine-grained variations within demographic groups or geographic areas. We recently used mobile device data to estimate the contact network in Portland, Oregon, and used it to simulate the epidemic across a population that had been partially vaccinated according to multiple strategies \cite{gulden2021protecting}. We found that targeted vaccination strategies that prioritize high-contact individuals far outperform a baseline strategy of vaccinating people at random for reducing cases, even when the targeted strategy is only imperfectly implemented.

In this work, we expand on that approach by performing a comparative study of the effectiveness of both social distancing and a vaccination strategy that targets those with the most physical contacts (such as workers in high-contact public-facing professions). This analysis is motivated by the fact that both mitigation strategies have significant challenges and high costs, so it is important to quantify their respective marginal benefits. Social distancing suppresses economic activity, has deleterious effects on child education, and leads to increased levels of depression and anxiety. Prioritizing the most high-contact individuals for vaccination has both direct challenges toward identifying those individuals and the opportunity costs of not prioritizing other vulnerable groups, and it could slow overall distribution. (We stress that the vaccination policy choice we are considering is not whether to prioritize vaccination, but whether to accept the costs associated with a \emph{targeted} vaccination strategy.) 
In this study, we only attempt to quantify the effectiveness of these two COVID-19 mitigation strategies for reducing the total number of cases; we do not model the number of hospitalizations or deaths, although these are also important factors for policy makers to consider. Nor do we consider other vaccination strategies, such as prioritizing the elderly, because our data source does not provide the necessary information to do so.

The benefits of these two strategies are not independent; the marginal benefit of a targeted vaccination strategy (compared to a uniform strategy) is strongly dependent on the contact network topology \cite{pastor2002immunization}, which in turn is directly and measurably affected by social distancing measures. By simulating the epidemic on empirical contact networks, both before and during social distancing, we are able to assess how the change in network topology affects the improvement of the targeted vaccination strategy over the uniform strategy. We also give a graph-theoretic explanation for the effectiveness of the targeted vaccination strategy, which is based on the asymptotic distribution of the node degrees in the contact networks. This analysis could be useful to policy makers for deciding the best courses of action in terms of both social distancing and the vaccination distribution strategy.

Previous modeling of optimal vaccine distribution strategies have used artificial contact networks derived from large-scale agent-based models \cite{chen2021prioritizing}, small real-world contact networks \cite{firth2020using}, and age-stratified ordinary differential equation models \cite{buckner2020dynamic, matrajt2020vaccine, bubar2021model}. Our approach is unique in that it uses a real-world contact network with more than 200,000 individuals.

\section{Empirical Analysis of Mobile Device Data for Portland, OR
\label{sec:network}
}
We were granted access to mobile device data for the city of Portland, Oregon by the software company UberMedia. This data consisted of approximately 2.2 billion device pings recorded between December 31, 2019 to April 13, 2020, and through it we were able to infer many of the physical contact interactions between individuals during this period, both before and after social distancing measures were enacted. (COVID-19 was officially declared a national emergency on March 13, 2020 by the President of the United States, and on March 23 Governor Kate Brown issued Executive Order 20-12, which instituted a number of social distancing policies that supplemented previous state-wide measures such as the closure of K-12 schools.\footnote{\url{https://www.oregon.gov/gov/admin/Pages/eo_20-12.aspx}} This section discusses the key features of these data.

\paragraph{Overall effect of social distancing on connectivity}
The mobile device data may be used to approximately infer when two people came into physical contact by detecting when they occupied the same location at the same time. Processing all 2.2 billion pings in this way results in a time series of person-to-person contacts that represented as a network with time-dependent edges. This network represents our estimate of the actual network across which COVID-19 spread, with the primary source of error being that not all interactions were represented in the raw data. In total, the graph consists of 13.5 million contacts between 327,363 unique devices.

\begin{figure*}
    \centering
    \includegraphics[width=1.0\linewidth]{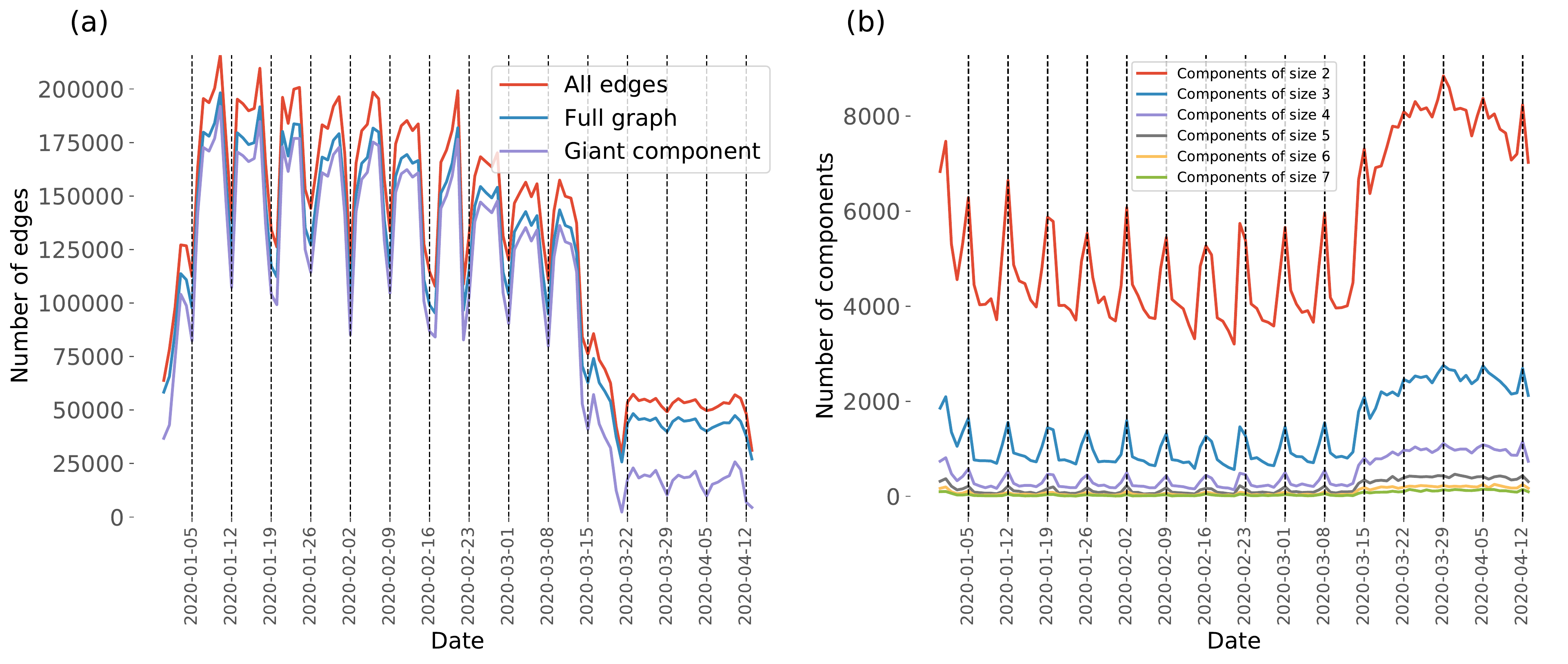}
    \caption{\textbf{Temporal variation of the contact network.} (a) The daily number of total contacts in the mobile device data, the daily number of edges in the temporal graph, and the daily number of edges in the giant component of the graph. The number of total contacts is slightly larger than the number of edges because some edges are duplicated because the two nodes came into contact multiple times during that day. The vertical dashed lines correspond to Sundays. (b) The number over time of small connected components containing between 2 and 7 nodes. The vertical dashed lines correspond to Sundays.}
    \label{fig:Gt_over_time}
\end{figure*}

Figure \ref{fig:Gt_over_time} (a) depicts  the number of daily contacts over the first months of the pandemic and shows how connectivity patterns changed over that period. There are weekly downward spikes due to standard reductions in mobility during weekends, and a clear and dramatic drop in connections beginning around mid-March. The plot also depicts the daily number of edges comprising the giant component (i.e. the largest connected component of the graph). The fraction of edges contained in the giant component also drops around mid-March. Figure \ref{fig:Gt_over_time} (b) shows the daily number of small components that contain between 2 through 7 nodes. Around mid-March the number of such small components increases, indicating that social distancing measures led to fragmentation of the network, potentially slowing the growth of the pandemic in Portland.

\paragraph{Change in person-to-person contact networks}
Due to data collection limitations, only a fraction of all real-world interactions between mobile devices were captured, and as a result the time-dependent graph has missing edges. We corrected for this under-counting by flattening the time series and creating a static weighted graph representing the superposition of all contacts, with each contact weighted according to its duration. In order to study the effect of social distancing measures, we separately applied this procedure to all contacts occurring before and after social distancing measures were enacted, and combined this with inferred data on home and work locations. The first network, $G_{\text{pre}}$, corresponds to contacts occurring between late February and early March of 2020, before any social distancing measures were adopted, and the second, $G_{\text{post}}$, captures contacts occurring between late March and early April, 2020, after social distancing measures were enacted. For each graph we discarded all but the largest connected component, so that the resulting network is connected. Appendix~\ref{sec:mobiledevice} 
contains the details of the procedure used to create these graphs from the raw mobile device data, and we have made the complete anonymized networks available here \url{https://github.com/RANDCorporation/network_vaccination}. A few salient features of the network topology demonstrate the effect of social distancing measures, and can also be used to understand network-based epidemiological models based on these graphs. Table~\ref{table:networksummary} lists key statistics of each network.\footnote{Although these networks are empirically derived from real data, they only capture a fraction of the nodes and edges of the true contact graph. The population of Portland, estimated at about 580,000 in the 2010 census, is larger than the size of either contact network, and the sampling method introduces some degree of sampling bias because the rates of smartphone ownership and participation in location tracking services vary across groups \cite{grantz2020use}. See \url{https://www.census.gov/quickfacts/fact/table/portlandcityoregon/PST045219}.}

\begin{table*}
\centering
   \caption{\label{table:networksummary} \textbf{Summary statistics for the two static contact networks derived from the time-dependent network.} Each network has been processed to discard all but the giant component. The density is defined as the fraction of possible edges that exist, i.e. $2M/N(N-1)$ for $N$ the number of nodes and $M$ the number of edges. $\langle \cdot \rangle$ denotes an average over all nodes in the network. $k$ is the node degree, $\kappa := \langle k^2 \rangle/\langle k \rangle$ is the heterogeneity parameter, $\langle \ell \rangle$ is the average shortest path length, and $\langle C \rangle$ is the average network clustering coefficient.}
\begin{tabular}{ccccccccc}
    \hline
    Network & Nodes & Edges & Density & $\langle k \rangle$ & $\langle k^2 \rangle$ & $\kappa$ & $\langle \ell \rangle$ & $\langle C \rangle$ \\ 
    \hline
    $G_{\text{pre}}$ (without distancing) & 214,393 & 1,538,092 & $6.69 \times 10^{-5}$ & 14.4 & 708 & 49.3 & 5.17 & 0.300 \\
    $G_{\text{post}}$ (with distancing) & 130,910 & 351,512 & $4.10 \times 10^{-5}$ & 5.37 &  88.2 & 16.4  & 7.41 & 0.323 \\    
    \hline
\end{tabular}
\end{table*}

\begin{figure}
    \centering
    \includegraphics[width=1\linewidth]{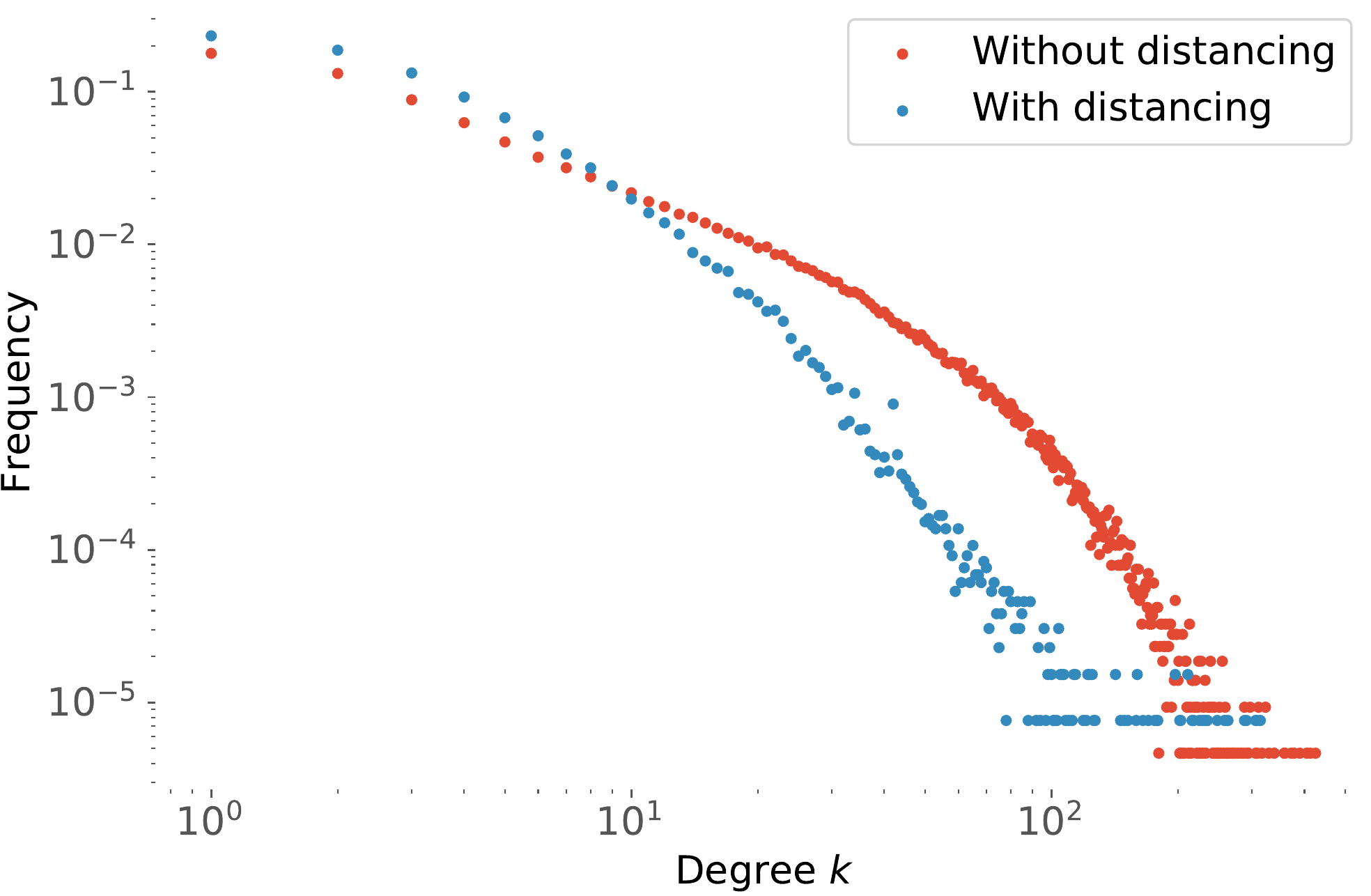}
    \caption{\textbf{The empirical degree distribution for the contact network both before and during social distancing.}}
    \label{fig:degree_dist}
\end{figure}

One of the most important network properties for percolation processes like the spread of a disease is the degree distribution. Figure \ref{fig:degree_dist} shows the degree distributions for both $G_{\text{pre}}$ and $G_{\text{post}}$. In both networks, the majority of nodes have just a few neighbors; for example, 65\% of the nodes in $G_{\text{pre}}$ and 88\% of those in $G_{\text{post}}$ have 10 or fewer neighbors. However, there are rare nodes with high degree in both plots. In 
Appendix \ref{sec:networkfit} 
we present the results of a statistical analysis of the tails of these distributions. We find strong statistical evidence that the degree distribution for $G_\text{post}$ is heavy-tailed rather than exponentially bounded. The statistics for the tail of the $G_{\text pre}$ degree distribution are less clear, but a heavy-tailed distribution is somewhat statistically favored over an exponentially bounded distribution. Dynamical processes on such heavy-tailed networks typically exhibit strong fluctuations, because although the high degree nodes are rare, their large number of ties to other nodes makes them very important hubs.

The heterogeneity parameter $\kappa := \langle k^2 \rangle/\langle k \rangle$ (where $\langle \cdot \rangle$ denotes an average over the degree distribution) provides another measure of the importance of fluctuations in a network's degree distribution: networks with $\kappa \gg \langle k \rangle$ are said to be heterogeneous. As two extreme examples, Erdős–Rényi graphs have $\kappa = 1$, and scale-free networks with degree distribution $P(k) \sim k^{-\alpha}$, with $2 < \alpha \leq 3$, have $\kappa = \infty$. The real-world contact networks are weakly heterogeneous, and social distancing can be seen to reduce heterogeneity, with $\kappa/\langle k \rangle \sim 3.5$ for the pre-social distancing network and $\kappa/\langle k \rangle \sim 3$ for the post social distancing network. There is empirical evidence of significant heterogeneity in the transmission of COVID-19, which has important implications for understanding its spread and the threshold for herd immunity \cite{adam2020clustering, wong2020evidence, hebert2020beyond, nielsen2021covid}. Heterogeneity in the transmission dynamics can be driven both by biological causes and by network effects, and real-world social networks such as these could serve as a useful tool for distinguishing the relevance of these two classes of effects for the heterogeneous transmission of COVID-19. 

One measure of the extent of public compliance with social distancing guidelines is the change in the tail of the degree distribution. Figure \ref{fig:degree_dist} shows that the social distancing policies had a dramatic impact on the number of high degree nodes, which significantly decreased relative to the pre-social distancing network as social distancing reduced the high-contact events such as going to gyms or dining in restaurants and bars. The magnitude of the effect is roughly an order of magnitude for $k \gtrsim 30$. It is important to note that these effects are not solely attributable to government-ordered social distancing measures - some social distancing can be expected to occur as a natural response to the worsening pandemic.

Statistical measures of ``closeness'' within the graph can further quantify the impact of social distancing on disease spread. One such measure is the average shortest path length $\langle \ell \rangle$ between any two random nodes. Social distancing increased $\langle \ell \rangle$ enough that an average of about two more intermediaries are required to connect an arbitrary pair of nodes. The empirical values of $\langle \ell \rangle$ are quite similar to the those for Erd\"{o}s-R\'{e}nyi random graphs with equal size and average node degree. Another measure of closeness is the average clustering coefficient $\langle C \rangle := \sum_{i=1}^N C_i/N$, where the local clustering $C_i$ of node $i$ equals the fraction of the possible links between the neighbors of node $i$ that are present. Interestingly, the socially distanced graph has a slightly higher clustering coefficient, which might be attributable to the development of small close-knit clusters (``pods'') during the pandemic. Taken together, these two results indicate that the real-world networks resemble small-world networks in that the average path length scales logarithmically with system size and there is a significant amount of clustering.

\paragraph{Theoretical impact of degree distribution on vaccination effectiveness}
The impact of the removal of high-degree nodes from the contact network greatly depends on the degree distribution. For networks with heavy-tailed degree distributions, such as scale-free networks, the high-degree nodes are extremely important: inoculating these nodes first greatly slows the spread of the disease, but a uniform vaccination strategy that misses these nodes has limited effectiveness. But for networks with exponentially bounded degree distributions, such as Erd\"{o}s-R\'{e}nyi or Watts-Strogatz graphs \cite{watts1998collective}, high-degree nodes are not much more connected than low-degree nodes, so the difference in effectiveness between targeted and uniform vaccination strategies is much smaller and in some cases negligible \cite{pastor2002immunization}.

A key question influencing COVID-19 vaccination implementation is therefore whether the real-world contact network is closer to a scale-free or an exponentially bounded network. The empirical contact networks are not strictly scale-free, although they do exhibit heavy-tailed degree distributions and a significant amount of heterogeneity. This suggests that these networks might be qualitatively similar to scale-free networks in that a targeted vaccination strategy would far outperform a uniform one.


\section{Simulating the Spread of the Epidemic over Empirical Contact Networks \label{sec:simresults}}
In order to directly test this hypothesis and to quantify the impact of social distancing on the spread of the epidemic, we separately simulated a Susceptible-Exposed-Infected-Removed (SEIR) model calibrated for COVID-19 on both the pre- and post-social distancing graphs $G_{\text{pre}}$ and $G_{\text{post}}$, both with and without vaccination. The details of the simulations may be found in 
Appendix \ref{sec:simdetails}.

\begin{figure}
    \centering
    \includegraphics[width=1\linewidth]{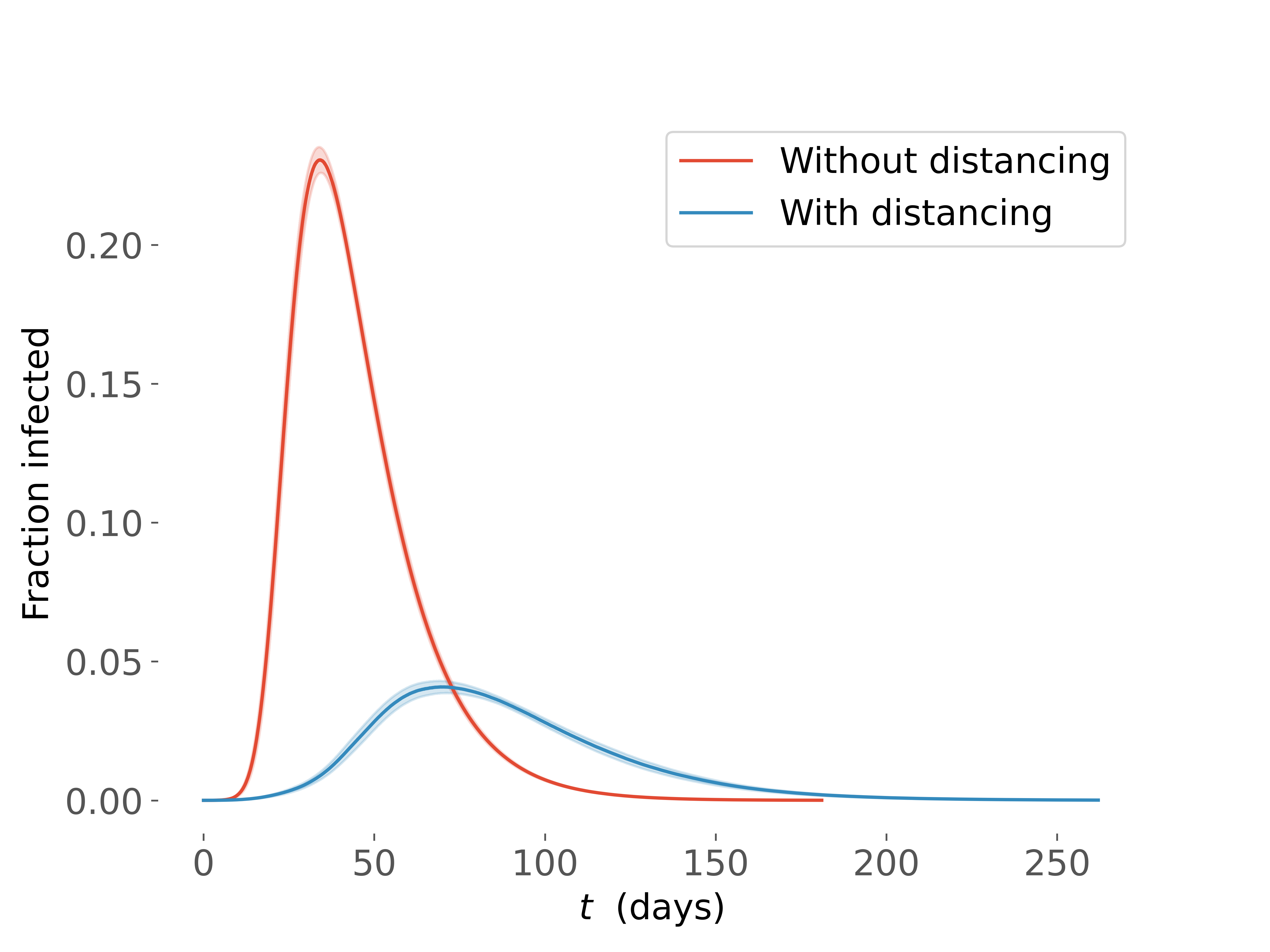}
    \caption{\textbf{The fraction of infected nodes as a function of time, for the contact networks with and without social distancing.} The shaded regions represent 95\% confidence intervals computed by averaging the results of 100 different runs.}
    \label{fig:SEIR_SD_comparison}
\end{figure}

\paragraph{Results}
First, we investigated the impact of the reduced connectivity caused by social distancing. Figure \ref{fig:SEIR_SD_comparison} depicts the trajectory of the epidemic when separately simulated on both networks, $G_{\text{pre}}$ and $G_{\text{post}}$. Social distancing is extremely effective at reducing both the cumulative impact of the epidemic and its peak intensity. The average number of total infections drops from 130,000 for $G_{\text{pre}}$ to 29,000 for $G_{\text{post}}$. This represents a 77\% decrease in total cases (or, taking into account the different sizes of the graphs, a 63\% decrease in the fraction of the population who became infected at some point). Similarly, the average peak number of infections drops from 51,000 for $G_{\text{pre}}$ to 6000 for $G_{\text{post}}$, corresponding to a 88\% decrease (or a 81\% decrease in the fraction of peak infections). This is a clear demonstration of the ``flattening of the curve'' goal that motivated the social distancing measures at the beginning of the pandemic. (In practice, social distancing is often implemented dynamically in response to a growing public health crisis, whereas here we have made the simplifying assumption that social distancing occurred from the very outset of the epidemic.)

\begin{figure*}
    \centering
    \includegraphics[width=0.8\linewidth]{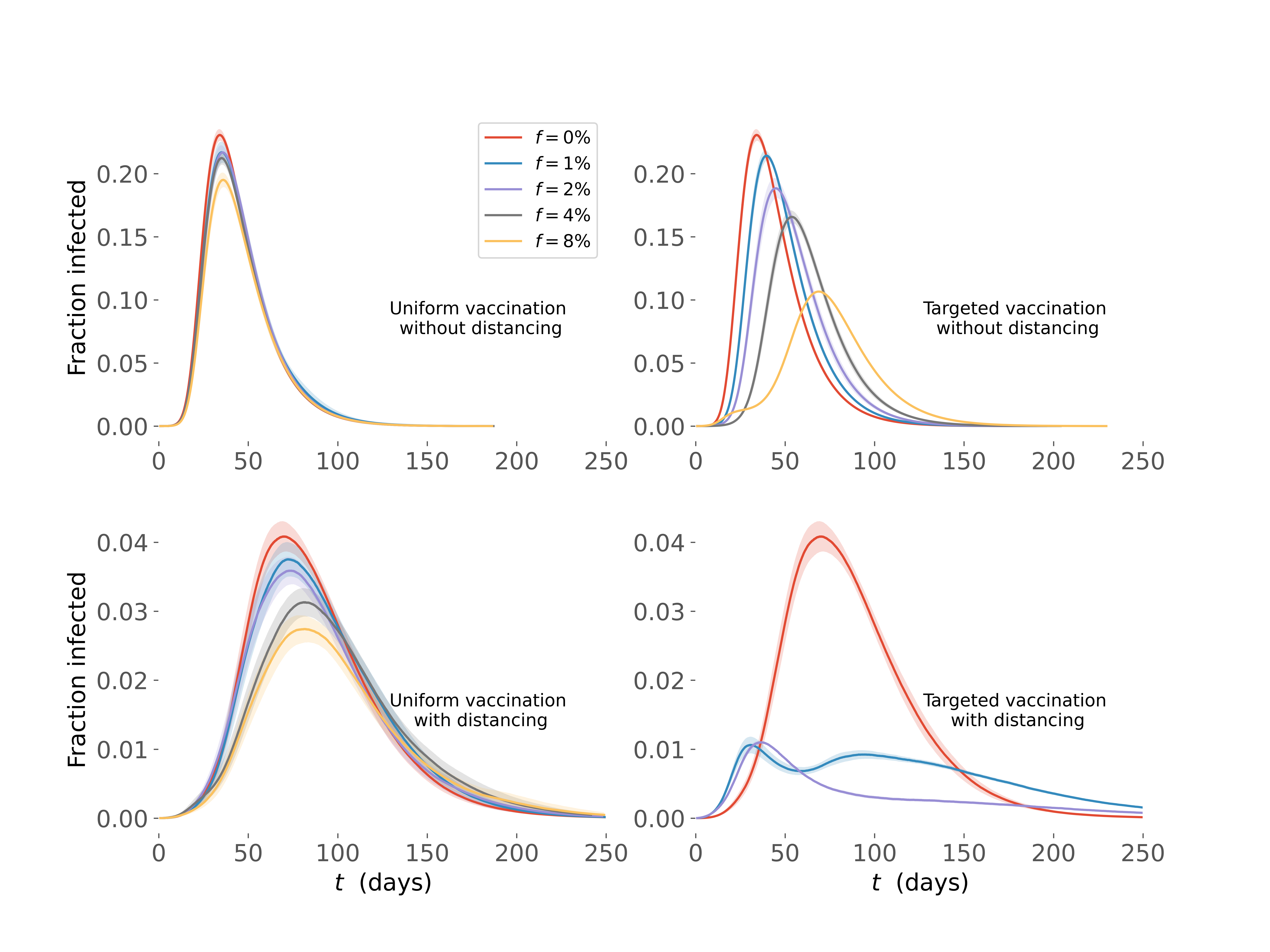}
    \caption{\textbf{The fraction of infected nodes as a function of time on networks without and with social distancing, and with varying fractions of the population vaccinated according to either the uniform or targeted strategy.} The shaded regions represent 95\% confidence intervals computed by averaging the results of 100 different runs. Note that the subplots have different vertical scales.}
    \label{fig:SEIR_SD_vacc_comparison}
\end{figure*}

Next, we investigated the impact of vaccinating a variable fraction of the population according to two vaccination distribution strategies, uniform and targeted. In both strategies a fraction $f$ of all nodes were granted perfect and instantaneous immunity, corresponding to simply removing them from the contact network. In the uniform strategy these nodes were chosen uniformly at random. In the targeted strategy, the top fraction $f$ of nodes, ranked by degree, were granted immunity. The effect of these strategies on the growth of the epidemic across both networks is shown in Figure \ref{fig:SEIR_SD_vacc_comparison}, which reveals that the targeted strategy is very effective at suppressing the growth of the epidemic on both networks.

\begin{figure}
    \centering
    \includegraphics[width=1.0\linewidth]{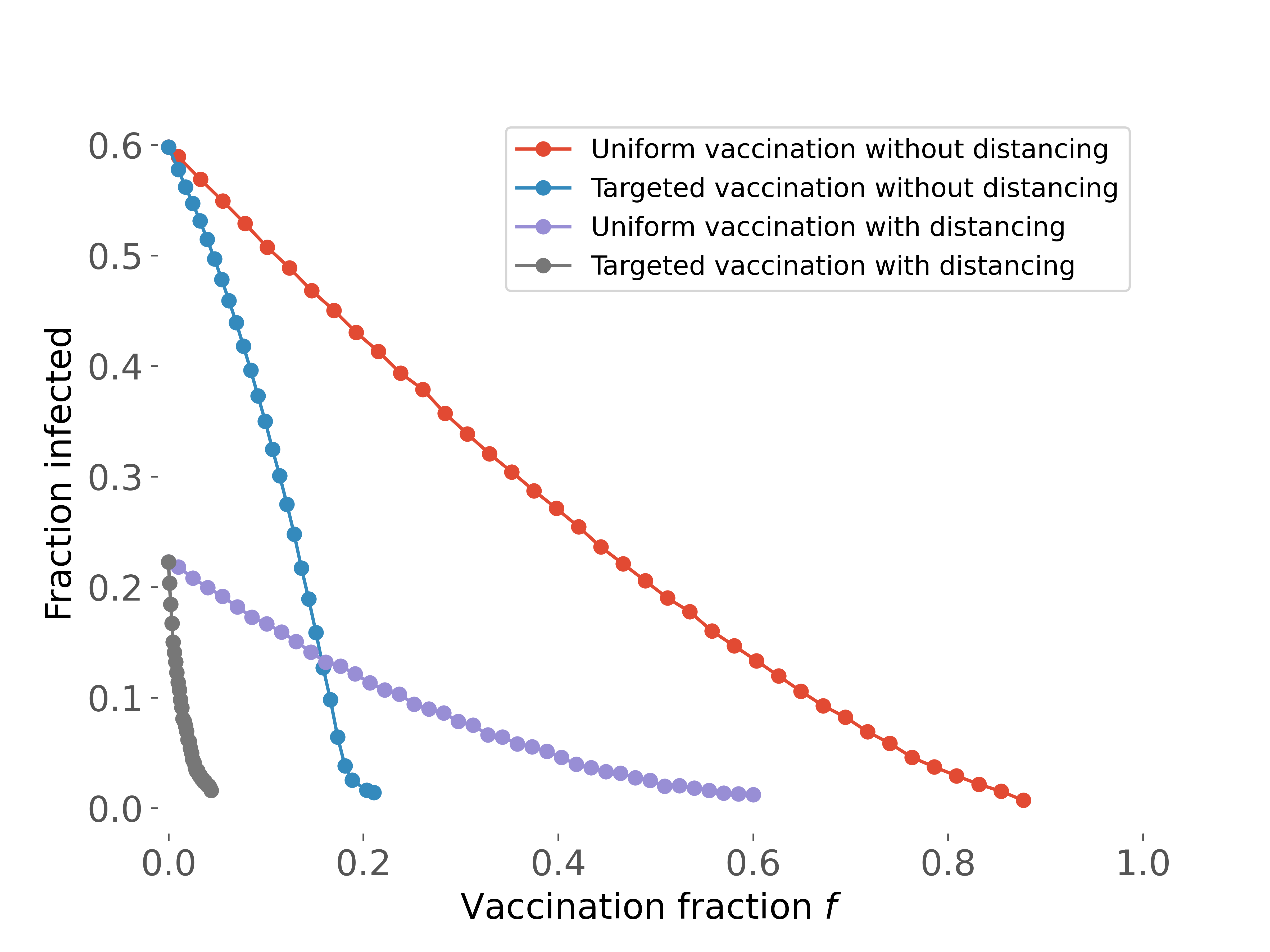}
    \caption{\textbf{The total number of infections caused by the epidemic in the presence of social distancing and/or vaccinations.} The cumulative fraction of the population which became infected at one point or other during the course of the epidemic is shown as a function of the fraction of vaccinated individuals $f$, for different social contact networks and in the presence of different vaccination distribution strategies. The targeted vaccine distribution strategy causes the total case count to quickly drop, and it clearly prevents the epidemic once the fraction of vaccinated individuals exceeds some threshold. In contrast, the uniform strategy suppresses the epidemic much more slowly and the total case count appears to approach 0 only as $f \rightarrow 1$.}
    \label{fig:SEIR_vacc_fraction}
\end{figure}

It is important for public health policy makers to understand the critical vaccination threshold required to prevent a disease from ever becoming endemic in a given community. (This threshold is related, but not identical, to the threshold for herd immunity in an ongoing epidemic, which we do not model.) Figure~\ref{fig:SEIR_vacc_fraction} depicts the cumulative fraction of infected individuals over the course of a simulation for a range of vaccination fractions. The critical vaccination threshold is defined to be the fraction $f_c$ above which the disease dies out without ever infecting an extensive fraction of the population. The threshold can be seen to be both highly dependent on whether social distancing is in place and on the distribution strategy for the vaccine. Regardless of social distancing, the targeted strategy far outperforms the uniform one and is able to suppress the epidemic when just a small fraction of the population has been vaccinated. The performance of the uniform strategy improves much more gradually as the fraction of vaccinated individuals increases. For the targeted strategies, the critical vaccination thresholds are estimated as $f_c = 0.23$ before social distancing and $f_c = 0.06$ during social distancing. Theory and empirical evidence suggest that the critical vaccine thresholds $f_c$ for both networks are formally equal to $1$, but Figure~\ref{fig:SEIR_vacc_fraction} shows that the infection rates become very low at intermediate levels of vaccination. 
Appendix~\ref{sec:vaxthreshold} 
further discusses our findings for the critical vaccination threshold.

\paragraph{Discussion}
The targeted vaccination strategy will always result in fewer infections than the uniform strategy, since it incorporates information about the person-to-person contact network. But as discussed earlier, the extent to which the targeted strategy outperforms a uniform strategy depends strongly on the network topology. Our simulation results show that the networks' degree distributions are heavy-tailed enough that the targeted strategy greatly outperforms the uniform one.

In our simulations, all the vaccinations occur at the outset of the epidemic. In reality, vaccinations against COVID-19 occur continuously throughout the epidemic, and it is likely that many of of the high-degree nodes will have been infected early on in the pandemic and will have therefore acquired immunity. This effect could similarly reduce the effectiveness of a targeted strategy. Furthermore, a fraction of these high-degree nodes with acquired immunity will have been asymptomatic, and therefore they may be redundantly vaccinated in a targeted strategy, which would present a problem if vaccine doses were severely limited. As both policies (targeted vaccination and social distancing) involve serious trade-offs and are difficult to implement, it is important to understand how the disease progresses in the presence of both.

Figure \ref{fig:SEIR_vacc_fraction} shows that the targeted strategy remains extremely effective even when social distancing has already reduced the connectivity: the combination of both policies very quickly prevents the spread of the epidemic. Indeed, targeted vaccination without social distancing is even more effective than social distancing and uniform vaccination. The gap between these two options is significant, with the epidemic being preempted after about 60\% and 20\% of the population has been immunized, respectively.

\paragraph{Limitations}
Our primary motivation in this work was not to construct a perfectly faithful COVID-19 model, but to study the role of real-world contact networks on the epidemic dynamics. As such, our models make several simplifying assumptions that do not hold in the real world.

First, the transition between disease states was taken to be Markovian, with transitions occurring at a constant rate. This is a reasonable assumption for the infection transition $S \rightarrow E$, but it is violated for the incubation $E \rightarrow I$ and recovery $I \rightarrow R$  transitions observed in the COVID-19 pandemic. Furthermore, we neglect biological heterogeneity in the transmission dynamics and assume a single homogeneous rate $\beta$ for all $S \rightarrow E$ transitions. This assumption removes the possibility for biologically-driven superspreading events, which other models have shown to significantly impact the dynamics of the epidemic \cite{nielsen2021covid}.

Second, our composite contact networks do not perfectly capture the true pattern of social contacts. The construction of the empirical contact network from raw mobile device data was complicated by data collection issues that are described in 
Appendix \ref{sec:mobiledevice}. 
We modeled the total population as closed and assumed that an external infection only entered the population once, which was probably not the case. In reality these networks are time-dependent, with edges existing only for finite periods. For methodological convenience, we instead used static graphs and represented the duration of contacts using edge weights. The extent to which epidemic dynamics on a static, weighted graph approximate epidemic dynamics on a time-dependent graph is unknown, and this is an important point which warrants investigation. Also, the high level of social distancing observed in the two weeks after the emergency declarations of mid-March 2020 may not have been sustained as the pandemic continued. Finally, our model does not incorporate the dynamic impact of infection on contacts, as many people limit their contacts while they were infected (although in the case of COVID-19, this effect may have been reduced by the prevalence of asymptomatic transmission).

Third, our model disseminates the vaccines all at once and before the epidemic begins, which is in stark contrast to how the vaccines have been rolled out during the ongoing COVID-19 pandemic. We also considered an idealized vaccine which is 100\% effective at both preventing the vaccinated node from becoming infected and at allowing that node to spread the disease to others. Recent studies have indicated that the COVID-19 vaccines are highly effective at preventing symptomatic infection and confer some protection against transmission as well  \cite{levine2021decreased, thompson2021interim}, but we do not yet have enough real-world data to assess the accuracy of this approximation.

\section{Conclusion}
We used mobile device data to measure the impact of social distancing measures on the network of person-to-person contact networks over which COVID-19 spreads. Social distancing results in fewer connections overall, with the greatest reduction occurring amongst those with many contacts. We then used these networks to simulate the spread of COVID-19 in order to assess the effect of the reduced connectivity. Unsurprisingly, the fewer connections in the socially distanced network greatly reduced the spread of the disease and resulted both in fewer total infections and in a reduced peak intensity of the epidemic. 

We also simulated the epidemic in the presence of two vaccination strategies, a uniform strategy where nodes are vaccinated at random, and a targeted strategy where nodes are selectively targeted for vaccination based on their degree. Our results show quite definitively that within the assumptions of our model, the targeted vaccination strategy is far more effective at reducing infections than the uniform strategy, even in the presence of social distancing that greatly reduces the overall number of high-contact individuals. (Our targeted vaccination strategy did not incorporate the duration of contacts, and we expect that doing so would further improve the strategy's efficacy.) We believe that the differences in the modeled effectiveness of various mitigation strategies are large enough to provide a proof of principle that this kind of modeling on empirical networks can be useful for health researchers and policy makers as they continue to weigh the costs and benefits of prolonged social distancing measures and implement national vaccination programs. (The question of how many resources to dedicate to vaccination targeting is still highly relevant for policy makers even in nations that have universal vaccine eligibility, because vaccination prioritization can also be implemented via targeted messaging, the locations of distribution sites, paid leave policies, etc.) Our results also provide motivation for further refining our modeling to more accurately capture aspects of the epidemic that were simplified, such as the dynamic nature of both the contact network and the vaccine rollout.

There are multiple related objectives that one might like to achieve with a disease mitigation strategy - for example, minimizing the total number of deaths, total number of infections, peak number of infections, duration for which the number of infections exceeds some threshold, disparate impacts on disadvantaged sub-populations, probability of emergence of virus variants, etc. For reasons of both data availability and modeling feasibility, we only considered a single objective function -- the total number of infections -- which serves as an imperfect proxy for the other objectives. For example, we had no way of determining individuals' demographic information from anonymized cell phone data, so we were unable to incorporate the variation in COVID-19 mortality rates between different demographic groups in order to directly estimate the different mitigation strategies' impacts on hospitalization and death rates. Which mitigation strategy to implement will depend both on one's choice of objectives and on insights gained from quantitative modeling.

Even after one has selected the outcomes to optimize, the implementation of any vaccination strategy will involve many practical considerations not captured by our model. First, vaccine hesitancy will ensure that not every eligible person will elect to receive the vaccine. Second, any network-based vaccination strategy will be difficult to implement in practice because it will require an imperfect estimation of the number of contacts each individual in the population has. These two effects can be expected to reduce the performance of the targeted strategy relative to the uniform one, but earlier work found that they do not qualitatively change the results, provided the sources of error are not too large \cite{gulden2021protecting}. Third, the practical challenges of prioritizing certain subgroups could slow down the overall vaccine distribution beyond the constraints due to limited supply. Finally, any network-based vaccination strategy runs the risk of creating a moral hazard effect where people are incentivized to manipulate their contacts in order to move ahead in the queue. A careful evaluation of these and many other considerations will be necessary for the planning and coordination of a vaccination distribution plan.

\subsection*{Data and Materials Availability}
The person-to-person contact networks used in this study were created by processing billions of mobile device pings in the greater Portland area over the course of multiple months. As described above, this raw data was then processed into contact networks which were in turn used in the epidemiological simulations. The raw data cannot be released for privacy reasons, but the weighted contact networks used in the simulations have been made available in the code repository. These networks are completely anonymized and do not contain any location or time-stamped information which could be used to possibly identify individuals. Furthermore, the code used the simulations and data analysis presented in this work has been made publicly available.\footnote{\url{https://github.com/RANDCorporation/network_vaccination}} The epidemic simulations made use of the publicly-available \textit{Python} library Epidemics on Networks \cite{miller2020eon}.

\subsection*{Author Contributions}
TG conceived of the initial idea and secured funding. TG, GSH, and RV designed the research. TG and DK processed the mobile device data to create the contact networks. GSH performed the network analysis, the epidemic modeling, and wrote the first draft of the manuscript. EP suggested methodological refinements to the SEIR modeling and helped to structure the presentation of the research results and conclusions. All authors discussed results and further edited the manuscript. All authors approved the final version.

\subsection*{Acknowledgements}
We thank UberMedia for graciously sharing with us the mobile device data that made this research possible, as well as our RAND colleagues Jeanne Ringel, Susan Marquis, and Anita Chandra for securing funding for this project. We also wish to thank Robert Axtell and Carter Price for reviewing an early version of this work. This research was conducted in the Community Health and Environmental Policy Program within RAND Social and Economic Well-Being, a division of the RAND Corporation. This work was partially funded by philanthropic gifts from RAND supporters and income from operations. 

\appendix 

\section{Inferring Contact Networks from Mobile Device Data \label{sec:mobiledevice}}
Mobile device data provided by the location intelligence company UberMedia was used to generate person-to-person physical contact networks for the city of Portland. 
Two separate contact networks were created, one corresponding to the period before widespread social distancing measures were adopted (late February and early March, 2020) and one corresponding to a period during social distancing (late March and early April, 2020). The networks' contact data was collected over equally long time periods.

The raw mobile device data consisted of roughly 2.2 billion pings from mobile devices (mostly smartphones) that were using one of about 150,000 apps that allow volunteered sharing of location data with the app developer and its partners. Each ping is associated with an anonymized device identifier and includes both the time and the geographic location of the device with a nominal precision of three meters. Because reporting from these devices is quite uneven, we developed a procedure to composite data from across about a month to build up a general movement pattern for about 250,000 of the roughly 1.7 million individuals in the Portland Metropolitan Area (PMA). We restricted the analysis to devices that had a common evening location within the PMA and appeared in the data on at least 10 days in both the pre-social distancing and post-distancing periods.

The process by which we converted pings into contacts involved two steps: first we identified the locations where each device stopped moving for some period of time, then we determined when devices were stopped near one another during overlapping times to produce contacts. We used four types of stops. Common Evening Locations (CELs) were estimated by UberMedia by identifying the location where each device was most often found between 7:00 p.m. and 7:00 a.m. Not every device had such a location and we excluded those that either did not have a CEL or where the CEL was outside of the PMA. Common Daytime Locations (CDLs), also estimated by UberMedia, reflected a similar location where a device was often found between 8:00 a.m. and 5:00 p.m. Devices that did not have a CDL were retained in the data set because many people do not have the kind of jobs that keep them in one location all day. We then identified devices that entered a radius of 10 meters from the center of a set of about 9,000 known business locations in Portland. If the device pinged within that business, we assumed that it was there for at least five minutes, though if it pinged repeatedly, we measured the length of the visit if it exceeded five minutes. Finally, we identified places where a device moved by less than 10 meters between successive pings that were temporally separated by less than an hour, thus identifying instances where devices stayed in one place for some duration. This had the effect of removing devices that were in motion (e.g. car navigation apps) and capturing activity taking place in a wide variety of locations (homes, parks, etc.).   

With these stops determined, we made some broad assumptions in order to compile them into a composite movement pattern that compensated for the fact that the volunteered mobile device data represented only a sample of actual activity. First, we assumed that each person spends 12 hours at home each day—from 8:00 p.m. to 8:00 a.m. Second, we assumed that each person who has a common daytime location spends eight hours at that location each day—from 9:00 a.m. to 5:00 p.m. Finally, we took about two weeks’ worth of the other stops and collapsed them down to a single day while retaining the time of day for each stop. We calibrated the number of days to use in a way that produced an average number of contacts per person that corresponded to activity estimates developed by the Network Dynamics and Simulations Science Laboratory at Virginia Tech \cite{maratheSyntheticDataProducts2014}.

Because the locations captured in the second two types of stops (business locations on the one hand and successive pings within 10 meters on the other), this method produced instances where the same person appeared to be in more than one place at the same time -- if, for instance, the device pinged outside of its common daytime location between 9 a.m. and 5 p.m., or if the device pinged at the same coffee shop on several successive days. This does not present a problem for purposes of understanding the frequency with which a person is likely to visit various locations, with more frequent locations being represented more often than less frequent ones.  The resulting set of stop locations should be thought of not as a history of movement for any given day but rather as a representative composite movement pattern.

Finally, we divided the PMA into 10-meter squares and looked for instances where two devices were in the same square during overlapping times.\footnote{Note that with this approach it is technically possible to have had two devices very close to one another and not have the interaction count as a contact because the devices happened to lie in neighboring squares. Stated another way, two devices can only be regarded as coming into contact if they are within $10 \sqrt{2}$ meters of one another (the maximum separation two devices can have and still lie in the same 10 meter$^2$ square), but the converse is not true; it is possible for two devices to be arbitrarily close to one another and not have that interaction count as a contact. A more accurate (but harder to implement) approach would have been to consider the devices to be in contact if they came within some threshold $\ell_2$ distance of one another.} These overlaps were aggregated upward to produce a list of potential contacts between people—instances where people stopped within 10 meters of one another or within the same establishment for some amount of time. This contact network is far from perfect. Many people are home for more or less than 12 hours each day. Not everyone works eight hours a day, and not all of that work happens between 9:00 a.m. and 5:00 p.m. People do not use their mobile devices at statistically random times – so the times when they appear in the data cannot be assumed to be a random sample of their activity. We have no way of knowing what people were doing when they stopped within 10 meters of one another, whether they were wearing masks, etc. However, the network does represent an empirically derived view of the whole movement pattern of a city. 

\begin{figure}
    \centering
    \includegraphics[width=0.48\textwidth]{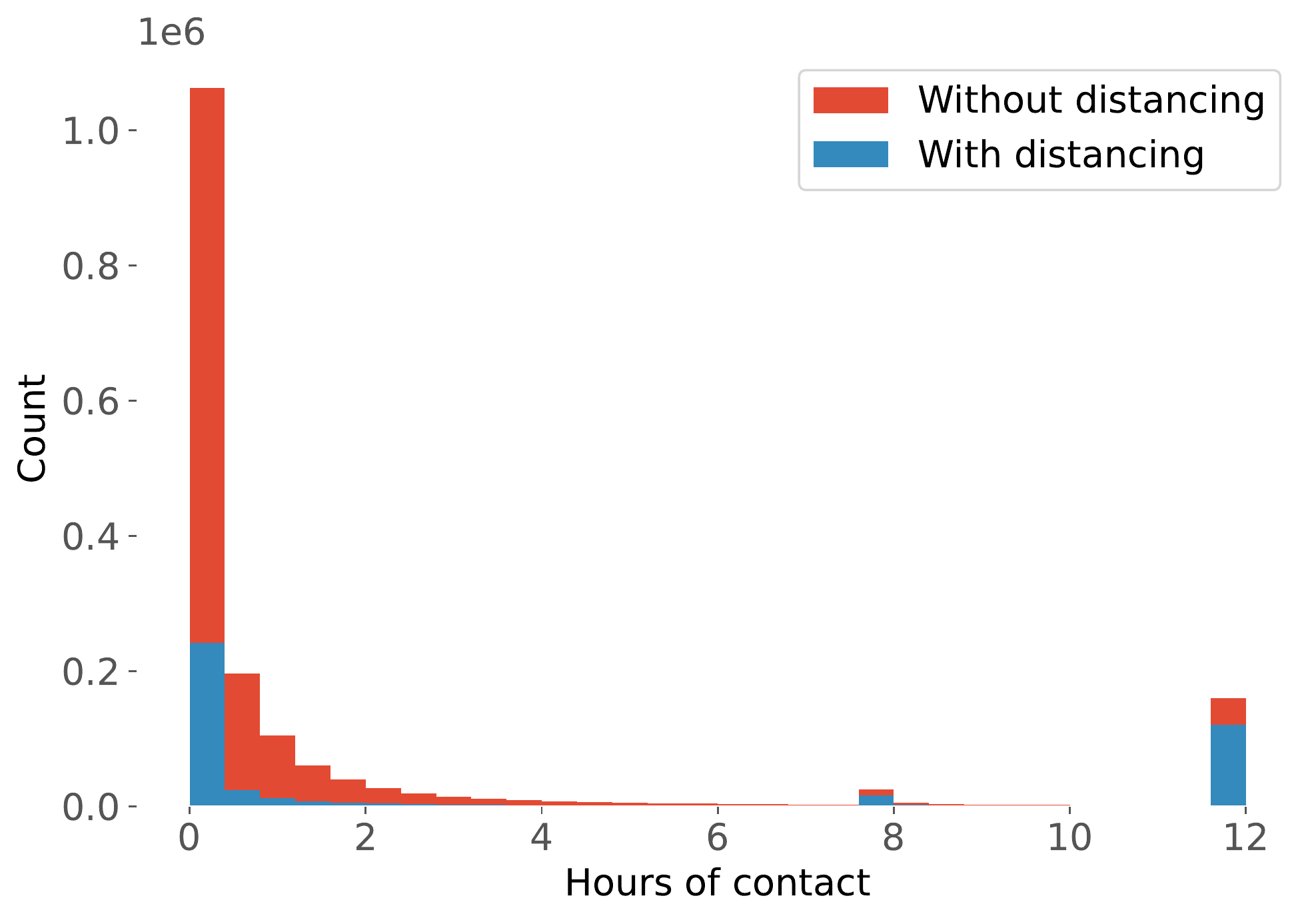}
    \caption{\textbf{The distribution of contact times in the raw mobile device data.} The peaks at 8 and 12 hours represent common daytime and nighttime locations, respectively.}
    \label{fig:contact_time_dist}
\end{figure}

After processing the 2.2 billion pings, we identified approximately 1.8 million person-to-person contacts in the period before social distancing measures were enacted, and 450,000 contacts after social distancing measures were enacted. The distribution of the durations of these contacts is shown in Figure \ref{fig:contact_time_dist}.

\section{Statistical Analysis of the Degree Distributions \label{sec:networkfit}}
Networks for which the tail of the degree distribution obeys a strict power law
are known as scale-free networks \cite{newman2018networks}. 
Many real-world networks across a large variety of disciplines have been claimed to be scale-free, although the actual prevalence of truly scale-free networks has been challenged \cite{broido2019scale}. In the context of epidemiological modeling, scale-free networks have been observed in the location-location contact network (as opposed to the person-person contact network) \cite{eubank2004modelling}. The extent to which the contact networks are scale-free can be assessed by performing a statistical fit to asymptotic power-law behavior \cite{clauset2009power}, which is described by a probability mass function of the form $p(k) = k^{-\alpha}$ for $k > k_{\text{min}}$, where $k_{\text{min}}$ indicates the start of the tail, and $\alpha > 1$ is the power-law exponent. We performed such a fit using the \textit{powerlaw} Python package \cite{alstott2014powerlaw}, which takes into account the various subtleties regarding the statistical significance of putative heavy-tailed behavior in finite data sets that was analyzed in \cite{clauset2009power}.

When we fit both parameters $(\alpha, k_{\text{min}})$ to the degree distribution of $G_{\text{pre}}$, the optimal $k_{\text{min}}$ value is 140, which implies that the tail only includes 0.4\% of the nodes, indicating that the fit should not be trusted. If instead of fitting $k_{\text{min}}$ we instead specify it to be the median, so that the tail contains 50\% of the nodes, then we find that the data is better described by a log-normal distribution than by a power-law, but both distributions are better fits than an exponentially bounded distribution. Thus, while not power-law, the tail is indeed described by a heavy-tailed distribution (though we note that there are various definitions of the term ``heavy-tailed'', and some of these exclude log-normal distributions). 

We then fit both parameters $(\alpha, k_{\text{min}})$ to the degree distribution of $G_{\text{post}}$. The optimal $k_{\text{min}}$ value is 21, which implies that the tail only includes 3.5\% of the nodes. This is a rather small proportion, although perhaps not small enough to discard the fit. Interestingly, this fraction is similar in scale to the critical vaccination fraction for the targeted vaccination strategy considered above, which indicates that the heavy-tailed regime includes enough of the high-connectivity nodes to determine the critical vaccination threshold. The fitted value of the exponent is $\alpha = 3.57$, and a pairwise comparison analysis concludes that the power-law fit better describes the data than a fit to either a log-normal distribution or an exponential distribution. We also considered a fit with $k_{\text{min}}$ again set to the median value, which, as in the case for $G_{\text{pre}}$, results in the fact that a log-normal distribution provides a better fit. 

Given the strong dependence of the fitting procedure on the choice for $k_{\text{min}}$, and the fact that allowing $k_{\text{min}}$ to be optimized over in addition to the exponent results in a rather short tail, the result of any one particular fit should be treated with some skepticism. However, the general conclusion that both distributions are heavy-tailed, with $G_{\text{pre}}$ perhaps only weakly so, seems to be robust to these subtleties.

Two common measures of ``small-world'' behavior in a graph are the average shortest path length and the average clustering coefficient. An Erd\"{o}s-R\'{e}nyi (ER) random graph with $N$ nodes and an average node degree $\langle k \rangle$ has an average shortest path length \cite{fronczak2004average}
\begin{equation}
    \langle \ell \rangle_\text{ER} = \frac{\ln N - \gamma}{\ln \langle k \rangle} + \frac{1}{2}
\end{equation}
(where $\gamma \approx 0.577$ is the Euler-Mascheroni constant), demonstrating the logarithmic growth in $N$ that is characteristic of a small-world graph. ER graphs with the same values of $N$ and $\langle k\rangle$ as our empirical graphs $G_\text{pre}$ and $G_\text{post}$ would have $\langle \ell \rangle = 4.89$ and $\langle \ell \rangle = 7.17$ respectively, which agrees well with the empirical values (5.17, 7.41, respectively). (It was computationally infeasible to average $\ell$ over all $N(N-1)/2$ pairs of nodes for each graph, which number in the tens of billions, and so we estimated $\langle \ell \rangle$ by averaging the shortest path length over $10^5$ randomly chosen pairs of nodes in each graph. All other statistics were averaged over the entire graphs.)

ER graphs have a much lower ($\Theta(N^{-1})$) average clustering coefficient than many empirical networks, which motivated the development of the Watts-Strogatz (WS) random graph model parameterized by $\langle k \rangle$ and the rewiring probability $p \in [0, 1]$, which has a similarly short average path length as the ER model but a $\Theta(N^0)$ average clustering coefficient \cite{barrat2000properties}
\begin{equation}
    \langle C \rangle_\text{WS} = \frac{3 (\langle k \rangle - 2)}{4 (\langle k \rangle - 1)} (1-p)^3.
\end{equation}
Fitting this model to the empirical graphs $G_\text{pre}$ and $G_\text{post}$ results in the fitting parameters $p = 0.244$ and $p = 0.176$ respectively.

\section{Simulation Details \label{sec:simdetails}}

\begin{figure*}
    \centering
    \includegraphics[width=0.48\textwidth]{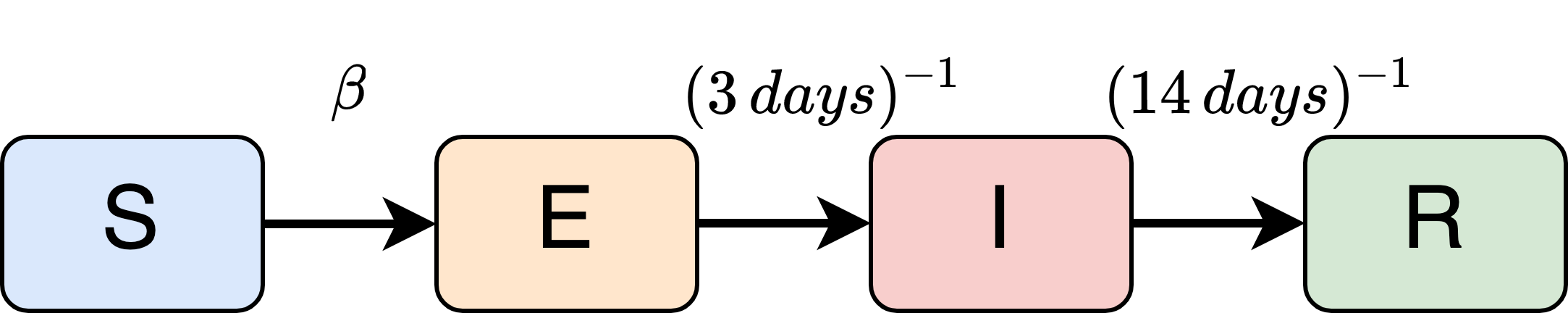}
    \caption{\textbf{The SEIR model dynamics.} Each transition occurs with constant rate. The $S \rightarrow E$ transition occurs only if the $S$ node is connected to one or more $I$ nodes in the graph, whereas the $E \rightarrow I$ and $I \rightarrow R$ transitions represent internal state changes.}
    \label{fig:SEIR}
\end{figure*}

\paragraph{Disease dynamics} 
The simulations model the epidemic at the individual level. Each node corresponds to a person whose condition with respect to the epidemic is represented in the node state, which can be $S$ (susceptible), $E$ (exposed), $I$ (infected), or $R$ (removed). The transitions between states are governed by the following rules. If an $S$ node is connected to an $I$ node, then there is a chance that the infection will spread to the $S$ node, which will cause its state to change to $E$. This transmission is modeled as a random event with constant rate $\beta$. Similarly, $E$ nodes will transition to $I$ with constant rate $a$, and $I$ nodes will transition to $R$ with constant rate $\gamma$. These dynamics are summarized in Figure \ref{fig:SEIR}. The $E\rightarrow I$ and $I \rightarrow R$ transitions represent internal state changes, whereas the $S\rightarrow I$ transition is mediated by a neighboring node in the network. 

The simulations were developed using the Epidemics on Networks (EoN) \textit{Python} library \cite{miller2020eon}. The constant rate transitions are simulated using the Gillespie algorithm  \cite{gillespie1977exact}, which is designed to exactly simulate continuous-time stochastic events that occur at a constant rate, without discretizing time. The algorithm exploits the fact that the time between transitions for a constant rate process is exponentially distributed, and therefore Markovian, in the sense that the probability of a transition occurring in the time interval $[t, t+ \Delta t]$ is independent of $t$.

\paragraph{Contact duration} 
The contact networks are represented by weighted graphs, where the weights correspond to the duration of the contact. This information was incorporated into the simulations by making the rate parameter for each edge proportional to the edge weight. The transmission rate for an edge $(i,j)$ in the SEIR model is taken to be $\beta_{i,j} = \beta w_{i,j}$, where $\beta$ is the global or base rate and $w_{i,j}$ is the weight. The weighted transmission rate may be given a simple interpretation by calculating the probability of a transmission event occurring within a time interval $\Delta T$, provided an $S$ node is connected to an $I$ node in the contact network: this probability is $p(\Delta T) = 1 - e^{-\beta w_{ij} {\Delta T}}$. For short contact times $p(\Delta T) \approx \beta w_{ij} \Delta T$, and so the weighting linearly scales the transmission probability. We note that this represents a crude approximation to the true way in which limited duration contacts affect the spread of COVID-19. The more correct approach would be to treat the edges in the graph as time-dependent, rather than as static but weighted. It is also worth noting that neither approach, treating the edges as time-dependent or as static and weighted, is able to capture the fact that not all person-to-person contacts are equivalently effective at spreading the disease. This would require incorporating important factors such as mask compliance, whether the interaction took place outdoors or indoors, the distance separating the two individuals during the interaction, and so on.

\paragraph{Model calibration} 
The SEIR dynamics are characterized by 3 rate parameters, $\beta$ (transmission), $a$ (incubation), and $\gamma$ (recovery). The time between state transitions is exponentially distributed with these rate parameters, so that the average incubation time is $a^{-1}$ and the average recovery time is $\gamma^{-1}$. In an attempt to match the real-world dynamics of COVID-19, these parameters were set to $\gamma^{-1} = 3$ days and $a^{-1} = 14$ days, respectively. The transmission rate $\beta$ was calibrated so as to cause the average number of infected to grow from 50 to 500 in 2 weeks, roughly matching what was observed in the course of the epidemic. Of course, social distancing was minimal during the first days of the epidemic, implying that the calibration should be done with respect with the network $G_{\text{pre}}$. Unfortunately, this resulted in a small $\beta$ that often caused the epidemic on $G_{\text{post}}$ to die out prematurely. We therefore performed the calibration with respect to the sparser network $G_{\text{post}}$.
This resulted in a $\beta$ value of $1.337 \, (\text{days})^{-1}$. The dynamics were made dimensionless by measuring time in units of days.

\begin{table}
\centering
   \caption{\label{table:parameters} \textbf{SEIR model parameters.} Each parameter describes the rate of the corresponding transition in units of (days)$^{-1}$.}
\begin{tabular}{cccccccc}
    \hline
    $\beta : S \rightarrow E$ & $a : E \rightarrow I$ & $\gamma : I \rightarrow R$ \\ 
    \hline
    1.337 & 1/3 & 1/14 \\
    \hline
\end{tabular}
\end{table}

\paragraph{Initial condition and model averaging} 
Each simulation initialized the epidemic by randomly selecting a degree-50 node to be infected, i.e. patient zero. This introduces a source of randomness into the simulations, in addition to the randomness due to the stochastic simulation of transition events for the nodes. To account for this, all results correspond to averages of 100 different runs. Due to fluctuations early in the simulation, some runs see the epidemic die out without infecting a large number of people. We therefore discarded runs which resulted in less than 1000 cases. If 10 or fewer runs out of a set of 100 runs survived this discarding process, then we discarded the entire set of runs. The disease trajectories in Figures~\ref{fig:SEIR_SD_comparison} and \ref{fig:SEIR_SD_vacc_comparison} 
depict the mean fraction infected at every point in time across all of the non-discarded runs, and the shaded regions represent 1.96 standard deviations about the mean (corresponding to a 95\% confidence interval if the variation between runs is normally distributed). One subtlety is that each simulation run corresponds to a distinct sequence of event times obtained using the Gillespie algorithm, and so the discrete trajectory sequences cannot be directly compared. To account for this, we used an interpolation to translate all trajectories to the same discrete set of grid points in time. 

\begin{figure*}
    \centering
    \includegraphics[width=0.8\textwidth]{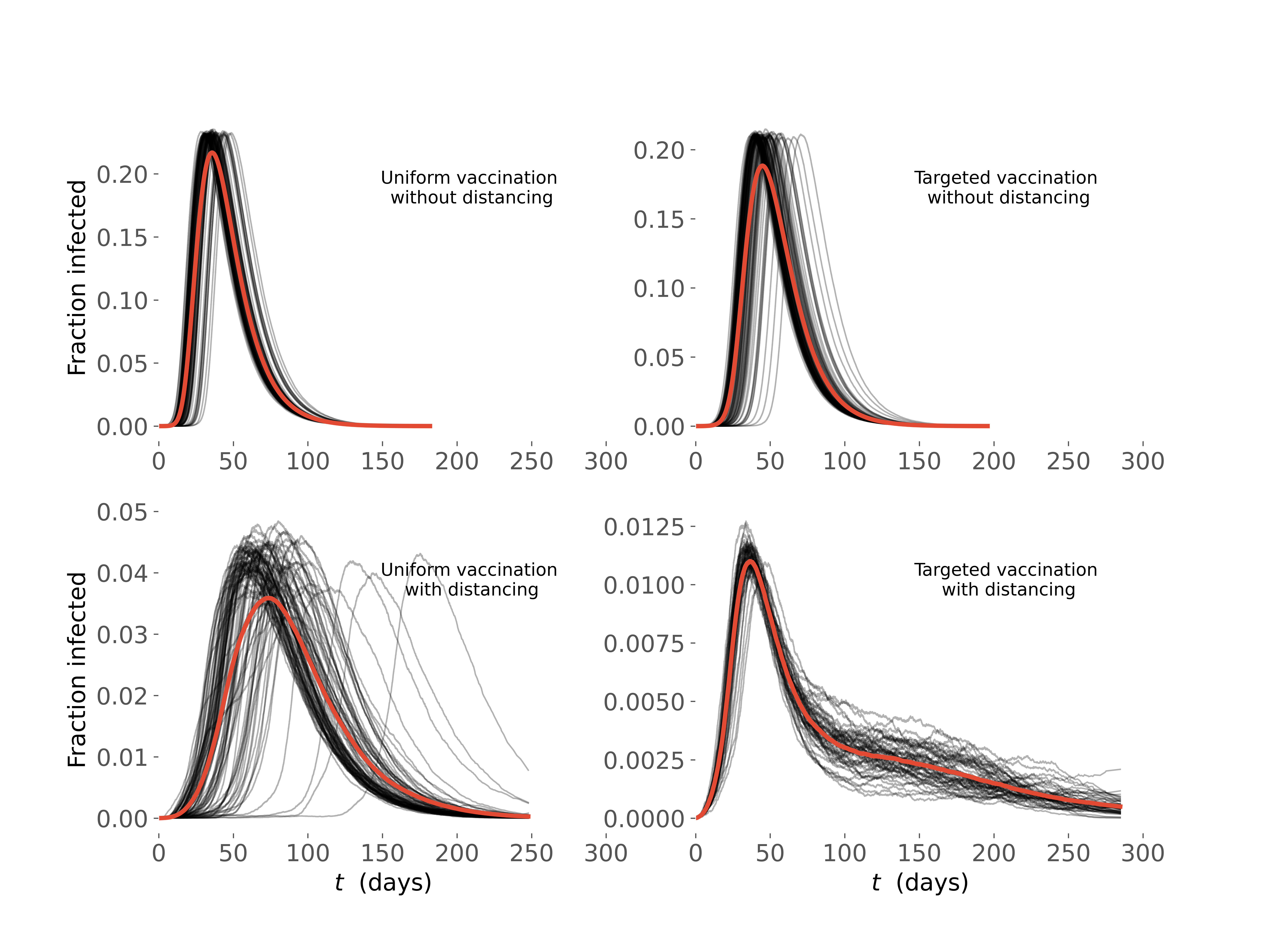}
    \caption{\textbf{Variability of epidemic trajectories.} The non-discarded epidemic trajectories (thin black lines) for a vaccination fraction value of $f=0.02$. The variability of the curves is due to different choices of initial conditions and the inherent stochasticity in the transmission process. The mean curve, obtained through the interpolation process described above, is shown in red.}
    \label{fig:SEIR_closer_look}
\end{figure*}

To give a sense for the variation between runs, in Figure \ref{fig:SEIR_closer_look} we depict all non-discarded runs for each of the four scenarios (with/without distancing and uniform/targeted vaccination) for a vaccination fraction of $f=0.02$. Evidently, although the peak intensity value is fairly constant between different runs, there can be significant variation in the time of peak intensity. Thus, averaging can lead to a shorter and broader mean curve than a typical trajectory. We note that the key quantities used to compare the contact networks and vaccination strategies, the peak infected fraction and total infected fraction, do not suffer from this effect.

\paragraph{Vaccination} 
We considered two vaccination strategies, uniform and targeted. In the uniform strategy, out of a population of $N$ nodes, a fraction $f$ is uniformly randomly selected to be vaccinated. In the targeted strategy, the nodes are sorted by their degree, and the top $N f$ nodes are selected for vaccination. We considered an idealized form of vaccination that happens instantaneously and before the epidemic begins, and we take the vaccine to be 100\% effective at preventing the vaccinated node from catching the disease and thus spreading it to others. From the perspective of the simulation, this simply corresponds to removing vaccinated nodes from the graph. 

\section{Critical Vaccination and Herd Immunity Thresholds \label{sec:vaxthreshold}}
In the context of an SEIR compartmental model on a family of contact networks with $N$ nodes, we define that a modeled population has achieved \textit{herd immunity} at a given time if a large enough fraction of the population is either vaccinated or recovered ($R$) that the extensive (i.e. $\Theta(N)$) contribution to the $I$ compartment is not increasing. If we define $i := \lim_{N \to \infty} I/N$ to be the intensive fraction of the population infected, then this means that $di/dt \leq 0$.

In many situations, a population has already been vaccinated against an infectious disease before the disease is introduced, with the goal of preventing the disease from ever becoming endemic in the population. In this context, it is reasonable to treat the vaccinated sub-population as static in modeling the spread of disease. For a given vaccination strategy, we define the \textit{critical vaccination threshold} $f_c$ to be the (static) minimum critical fraction of the population that needs to be vaccinated so that $I(t)$ scales subextensively with $N$ for all $t$ after the disease is introduced in a finite number of nodes. (In this case, in the limit of a large population size, only a negligible fraction of the population will ever be infected.) If the vaccinated fraction of the population exceeds the critical threshold, then herd immunity is immediately achieved at $t = 0$ because $i(t) \equiv 0$. Intuitively, the critical threshold describes the level of vaccination required to prevent the disease from ever becoming established at all, and not the level required to suppress an existing epidemic (which would depend on $R$). The critical vaccination threshold is therefore essentially a static concept (although in practice, numerical modeling needs to account for sub-extensive dynamics).

By contrast, if less than the critical threshold of the population is vaccinated when the disease is introduced into the population, then the infected compartment $I$ will grow extensively until the fraction of nodes that either become newly vaccinated or enter the $R$ state has grown large enough to make up the difference. (This is necessarily the case in a situation like the ongoing COVID-19 pandemic, in which no vaccines were available until a significant fraction of the population had already been infected.) In these situations, herd immunity is first achieved at some time $t > 0$, so it is intrinsically a dynamic concept. 

Our modeling made the simplifying assumption that all vaccines have already been administered when the disease is introduced into the population. We only attempted to calculate the critical vaccination threshold $f_c$ that prevents $I(t)$ from ever scaling extensively; calculating the actual herd immunity threshold for vaccination fractions $f < f_c$ would require also determining the time $t$ at which $i(t)$ peaks and begins decreasing.

Ref.~\cite{pastor2002immunization} derived analytic expressions for the critical vaccination thresholds for a simpler SIS model on both a Watts-Strogatz (WS) random contact network (whose degree distribution is exponentially bounded) and a scale-free Barab\'{a}si-Albert (BA) random contact network (whose degree distribution decays as a power law). The authors found that for either the uniform or the targeted vaccination strategy on the WS network, the equilibrium fraction of the population infected $i_\text{eq}$ vanishes linearly with the population's vaccinated fraction $f$: ${i_{eq} \propto f_c - f}$ for ${f \leq f_c}$. The critical vaccination threshold $f_c$ is the same for both strategies: the WS network is homogeneous enough that the highest-connectivity nodes are not much more connected than the average node, so the order of vaccination prioritization is not very important.

By contrast, the uniform vaccination strategy on the scale-free BA network leads to a positive equilibrium infection prevalence $i_\text{eq}$ for any vaccination fraction below $f_c = 1$. Near complete vaccination, the equilibrium infection prevalence scales as
\begin{equation}
i_\text{eq} \simeq 2 \exp[-c/(1-f)], \label{eq:targetedSF}
\end{equation}
where $c$ is a positive constant that depends on the contact network and infection spread modeling parameters. (However, the critical threshold only approaches $1$ logarithmically with increasing network size $N$, so it can be appreciably below $1$ for a finite network. Moreover, the exponentially fast decay of $i_\text{eq}$ near $f \simeq 1$ means that the infected population can become very small well before true herd immunity is reached.) On the other hand, the targeted vaccination strategy on the BA network does have a critical vaccination threshold $f_c \simeq \exp(-2c)$ that is strictly below $1$, and for this strategy the equilibrium infection prevalence $i_\text{eq}$ vanishes linearly as $f \to f_c$, as with the WS contact network.

This prior theoretical analysis was conducted for a simpler SIS model than the SEIR model that we used to model COVID-19. As discussed above, the notion of herd immunity is more subtle in the SEIR case, because both the vaccinated nodes and those that dynamically enter into the $R$ state contribute to herd immunity. Moreover, an SEIR model cannot demonstrate a positive equilibrium infection rate if the disease is only introduced into the population once, because eventually a critical fraction of nodes will enter the $R$ state. We therefore considered the cumulative number of infected nodes rather than the equilibrium fraction, and we assumed that scaling behavior similar to that derived in Ref.~\cite{pastor2002immunization} also applies in our more complex model.

Another complication is that Ref.~\cite{pastor2002immunization} only considered contact networks whose degree distributions are either exponentially bounded or power-law (i.e. scale-free). But the degree distributions of our empirical contact networks appear to be intermediately heavy-tailed -- heavier than exponential but less heavy than power-law, with a log-normal distribution showing the best statistical fit. It is therefore not clear whether our empirical networks would demonstrate infection behavior closer to the WS or the BA network model, so we investigated this question empirically.

For the targeted vaccination strategy, the theoretical infection prevalence vanishes linearly for both network topologies, so we performed a linear fit to the last few data points in 
Figure~\ref{fig:SEIR_vacc_fraction} 
to determine the critical vaccination thresholds. For the uniform strategy, the theoretical infection prevalence vanishes linearly for the WS topology, but Equation \eqref{eq:targetedSF} shows that it vanishes non-analytically at the critical threshold $f_c = 1$ for the BA topology, so a polynomial fit is not accurate. Moreover, it is computationally challenging to model the disease at high vaccination levels, because there are few nodes left susceptible.

\begin{figure}
    \centering
    \includegraphics[width=0.48\textwidth]{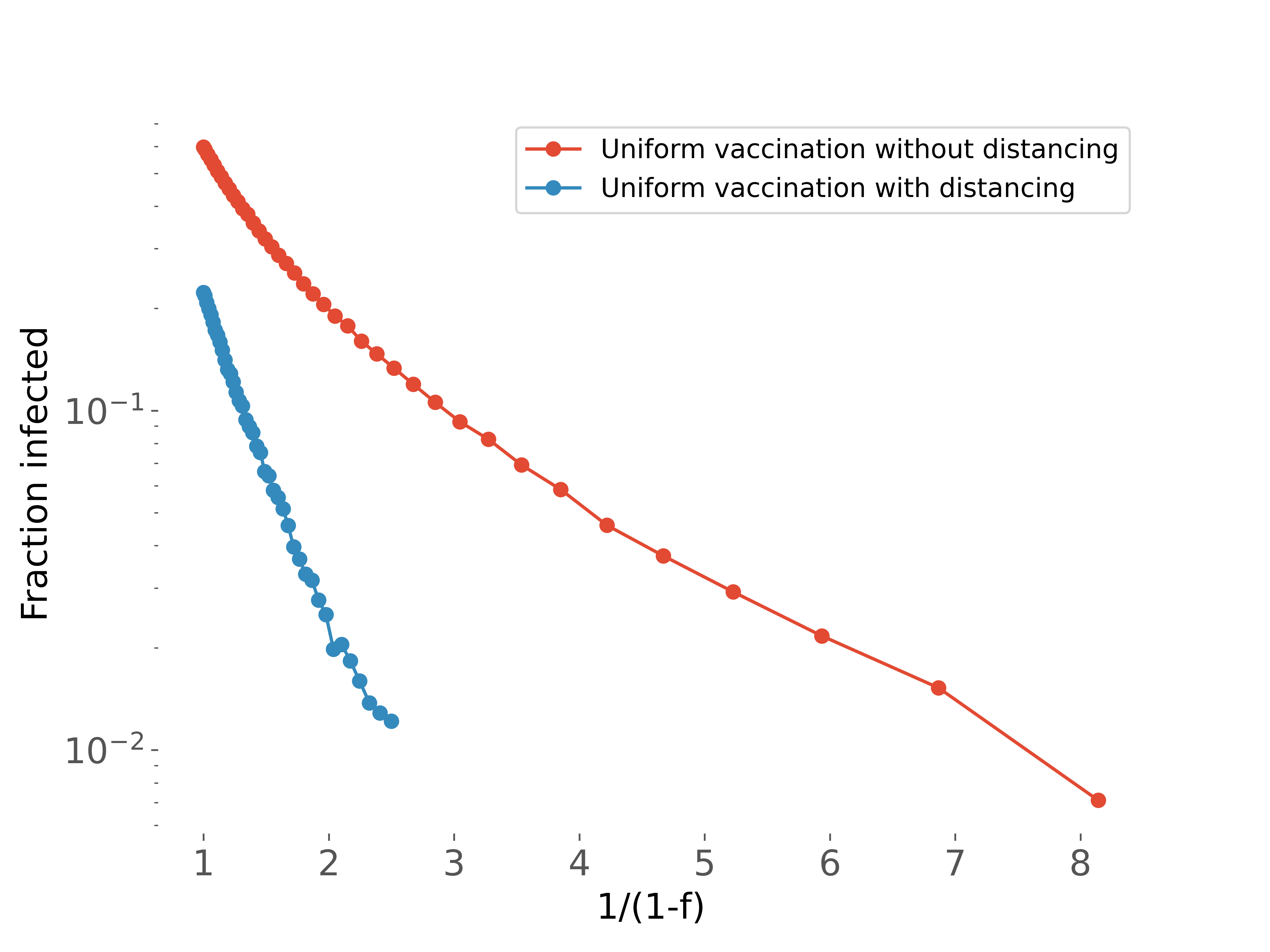}
    \caption{\textbf{Understanding the approach to a fully vaccinated population ($f=1$).} The modeling results from 
    Figure~\ref{fig:SEIR_vacc_fraction} 
    for the uniform strategies, with the horizontal axis remapped to $1/(1-f)$ and the vertical axis displayed on a logarithmic scale. Equation~\eqref{eq:targetedSF} predicts that the curves for the SIS model on a scale-free BA network eventually become straight when plotted in this way. The results from our SEIR modeling on empirical networks also appear quite straight, indicating that our model has the same critical vaccination threshold and scaling behavior.}
    \label{fig:log_reciprocal_plot}
\end{figure}

We therefore directly compared our data to the analytic form predicted by Equation \eqref{eq:targetedSF}, by taking the simulation results for the uniform strategies displayed in Figure~\ref{fig:SEIR_vacc_fraction} 
and replotting them in Figure~\ref{fig:log_reciprocal_plot} with a horizontal axis of $1/(1-f)$ and a vertical axis that displays the fraction infected on a logarithmic scale. We see that the curves eventually become quite straight when plotted in this way, which is what is predicted by Equation~\eqref{eq:targetedSF}. If the critical threshold $f_c$ were less than $1$, as with the WS topology, then the curves in Figure~\ref{fig:log_reciprocal_plot} would diverge to $-\infty$ at that value, which does not appear to be the case. We therefore conclude that the critical vaccination threshold $f_c = 1$ and the scaling behavior of the infection fraction for our model both appear to agree with the behavior of the SIS model on the scale-free BA network. This provides further evidence that with regards to the impact of vaccination, our empirical contact networks are better modeled as having heavy-tailed rather than exponentially bounded degree distributions.

\bibliography{refs}
\bibliographystyle{JHEP}

\end{document}